\documentclass[twocolumn,prl,superscriptaddress,floatfix]{revtex4}
\usepackage{graphicx}
\usepackage{xcolor}
\usepackage{epsfig}
\usepackage{rotating}
\usepackage{amsmath}
\usepackage{amsfonts}
\usepackage{amssymb}
\usepackage{enumerate}
\usepackage{longtable}
\usepackage{appendix}
\usepackage{dcolumn}% Align table columns on decimal point
\usepackage{bm}
\DeclareMathOperator{\Tr}{Tr}
\usepackage{hyperref}
\usepackage{braket}

\emergencystretch=\maxdimen
\begin{document}

\title{Superconductivity and charge density wave order in the 2D Holstein model}

\author{Owen Bradley}
\affiliation{Department of Physics, University of California Davis, CA 95616, USA}

\author{George G. Batrouni}
\affiliation{Universit\'e C\^ote d'Azur, CNRS, INPHYNI, Nice, France}
\affiliation{Centre for Quantum Technologies, National University of
  Singapore, 2 Science Drive 3, 117542 Singapore}
\affiliation{Department of Physics, National University of Singapore, 2
  Science Drive 3, 117542 Singapore}
\affiliation{Beijing Computational Science Research Center, Beijing
  100193, China}

\author{Richard T. Scalettar}
\affiliation{Department of Physics, University of California Davis, CA 95616, USA}

\date{\rm\today}

\begin{abstract}
The Holstein Hamiltonian describes fermions hopping on a lattice and
interacting locally with dispersionless phonon degrees of freedom.  In
the low density limit, dressed quasiparticles, polarons and
bipolarons, propagate with an effective mass.  At higher densities,
pairs can condense into a low temperature superconducting phase and,
at or near commensurate filling on a bipartite lattice, to charge
density wave (CDW) order.  CDW formation breaks a discrete symmetry
and hence occurs via a second order (Ising) transition, and therefore
at a finite $T_{\rm cdw}$ in two dimensions.  Quantum Monte Carlo
calculations have determined $T_{\rm cdw}$ for a variety of
geometries, including square, honeycomb, and Lieb lattices.  The
superconducting transition, on the other hand, in $d=2$ is in the
Kosterlitz-Thouless (KT) universality class, and is much less well
characterized.  In this paper we determine $T_{\rm sc}$ for the square
lattice, for several values of the density $\rho$ and phonon frequency
$\omega_0$.  We find that quasi-long range order sets in at $T_{\rm sc} \lesssim t/20$, where $t$ is the near neighbor hopping amplitude, consistent with previous rough
estimates from simulations which only extrapolated to the temperatures
we reach from considerably higher $T$.  We also show evidence for a
discontinuous evolution of the density as the CDW transition is
approached at half-filling.
\end{abstract}

\maketitle

\section{I. Introduction}

The interactions of electrons with lattice degrees of freedom
(phonons) underlie many of the fundamental properties of solid state
materials.  The many-body nature of the problem, however, poses
significant challenges to analytic investigation.  Hence, over the
last several decades, increasingly sophisticated computational methods
have been exploited to gain quantitative insight.  Early quantum Monte
Carlo (QMC) work on electron-phonon models focused on the dilute
limit.  As an electron moves through a material, the polarization of
the underlying medium causes a cloud of phonons to follow.
Simulations studied the resulting ``single electron polaron",
identifying its size and effective mass as functions of the
electron-phonon coupling and phonon frequency
\cite{kornilovitch98,kornilovitch99,alexandrov00,hohenadler04,ku02,spencer05,macridin04,romero99,bonca99}.
If the interaction is sufficiently large, it was shown that it is
possible for two polarons to pair.  The size, dispersion, and
stability of the resulting bipolarons was evaluated
\cite{hohenadler07,hague09,davenport12}, as well as bipolaron physics
across a range of fillings \cite{li20}.

As the density of these dressed quasiparticles increases, they can
condense into phases with long range order (LRO). One possibility is off-diagonal quasi-long range order, i.e.~superconductivity (SC).  At,
and close to, special commensurate densities, on a bipartite lattice,
diagonal LRO, i.e.~charge density wave (CDW) states, are another
possibility.  The competition between these two low temperature phases
is a fundamental feature of both materials
\cite{gruner88,gorkov89,gabovich10} and of simplified models of the
electron-phonon interaction.

One such model is the Holstein Hamiltonian \cite{holstein59}, which
describes electrons hopping on a lattice and interacting locally with
dispersionless phonon degrees of freedom. At commensurate filling on
bipartite lattices, it exhibits a transition to CDW order at a finite
$T_{\rm cdw}$ in two dimensions.  Early QMC studies of the Holstein
model examined the competition between CDW and SC on square lattices of up to $8 \times 8$ sites, observing the
enhancement of SC correlations and a simultaneous reduction in the CDW
structure factor as the system is doped away from half-filling
\cite{scalettar89, noack91, vekic92}.  Early estimates of $T_{\rm
  cdw}$ were obtained using a finite-size scaling approach, although
computational constraints on lattice size limited their accuracy.

The SC transition believed to occur away from half-filling (in two
dimensions) belongs to the Kosterlitz-Thouless (KT) universality
class.  Although similar attempts were made to quantify its
appearance, it remains much less well characterized.  Veki\'{c} et.~al
\cite{vekic92} provided estimates for $T_{\rm sc}$ based on a
finite-size scaling of QMC data for the same lattices of up to $8
\times 8$ sites, as were analyzed for the CDW transition, but only
reached inverse temperatures $\beta \leq 12/t$.  The computational
limitations on both temperature and lattice size which restricted
simulations to these ranges prevented an accurate finite-size scaling
to be performed.  For phonon frequencies $\omega_0/t = 1$, it was
estimated that the SC transition occurs within an approximate range
$\beta_{\rm sc} = 30 \textrm{--} 40$, more than a factor of two colder
than the lowest temperatures simulated. Finite-size scaling estimates
of the critical temperature at higher phonon frequencies, which would
tend to have higher, and hence more accessible, $T_{\rm sc}$ were also
limited in accuracy.

More recent studies of the Holstein model have refined estimates of
$T_{\rm cdw}$ at half-filling on the square lattice \cite{costa17,
  weber18, costa18}, and studied the interplay between SC and CDW
order as electron-phonon coupling is varied \cite{Sykora_2009}.  The
influence of phonon dispersion on both SC and CDW ordering has also
been studied \cite{costa18}, with strong evidence found for the onset
of SC at half-filling when phonon dispersion is present. A finite-size
scaling analysis obtained $T_{\rm sc} \approx t/26$ at a phonon
frequency $\omega_0/t = 4$, simulating lattices of up to $12 \times
12$ sites. Recently, the CDW transition in the Holstein model has also
been investigated for both the honeycomb and $\pi$-flux geometries
\cite{zhang19, feng20, zhang20}, as well as for the square lattice
with anisotropic hopping amplitudes \cite{cohenstead19}.  These
studies focused on the half-filled case only and hence did not advance
our understanding of $T_{\rm sc}$.  Recent work on the triangular
lattice Holstein model \cite{li19} has shown that frustrating the
charge order via a non-bipartite lattice can enhance SC, and an
estimate of $T_{\rm sc} \approx t/10$ was obtained at a phonon
frequency $\hbar\omega /E_F = 0.3$ (where $E_F$ is the Fermi
energy). This estimate was obtained at half-filling through a
finite-size scaling analysis, using lattices up to $12\times12$ sites.
However, in the work of \cite{li19}, no analogous evidence of the SC
transition was observed for the square lattice for the parameters
studied.

In the present paper, we resolve this situation by determining $T_{\rm sc}$ for the square lattice for several values of the phonon
frequency $\omega_0$ and electron density $\rho$ away from
half-filling.  We perform QMC simulations of lattices up to $12 \times
12$ sites, at inverse temperatures up to $\beta = 28/t$.  Through a
finite-size scaling analysis we find that SC sets
in close to, but still above, the lowest temperatures simulated.  That
is, our study does not rely on an extrapolation from temperatures much
higher than $T_{\rm sc}$.  We also investigate the variation of the
CDW structure factor with wave vector as the system is doped away from
half-filling, finding evidence for a possible incommensurate CDW phase
at low temperature.

We note that, in addition to the computational literature cited above,
considerable effort has gone into the analytic solution of the
Holstein Hamiltonian.  The Migdal-Eliashberg (ME) equations
\cite{migdal58,eliashberg60} form the basis for much of the analytic
work on strongly coupled electron-phonon models, but disagree with
exact QMC simulations
\cite{scalettar89,noack91,vekic92,niyaz93,noack93}, especially as the
temperature is lowered at densities in the vicinity of half-filling
where competing CDW formation occurs.  This comparison can be improved
somewhat with `renormalized ME' theory in which the phonon propagator
is dressed by electron-hole bubbles \cite{marsiglio90}. Recently,
there has been renewed interest in examining the limits of ME theory
and when it breaks down
\cite{esterlis18,mishchenko20,chubukov20,dee19,dee20}. Indeed, it has
been shown that ME can work well for $\omega_0 << E_F$ provided the
electron phonon coupling is not too large, enabling estimates of
$T_{SC}$ to be made by extrapolating DQMC results down to lower
temperatures using ME calculations \cite{esterlis18}. However, we note
that several of the parameter sets we study in this work are outside
the limits of ME theory.

\section{II. Model and Methods}

The Holstein model is a tight-binding Hamiltonian which describes the
interaction between electrons and local phonon modes in a lattice
\cite{holstein59}, 
\begin{multline}
\label{ham}
\hat{H} = -t \sum_{\langle \mathbf{i},\mathbf{j} \rangle, \sigma} \left(
\hat{c}^\dagger_{\mathbf{i}\sigma}\hat{c}^{\phantom{\dagger}}_{\mathbf{j}\sigma} + h.c.\right)
 - \mu \sum_{\mathbf{i} \sigma}\hat{n}_{\mathbf{i}\sigma} \\
+ \frac{1}{2}\sum_{\mathbf{i}} \hat{P}_\mathbf{i}^2 +
\frac{\omega_0^2}{2}\sum_\mathbf{i}\hat{X}_\mathbf{i}^2 
+ \lambda\sum_{\mathbf{i} \sigma} \hat{n}_{\mathbf{i}\sigma} \hat{X}_\mathbf{i} \,\,.
\end{multline}
Here $\hat{c}^\dagger_{\mathbf{i}\sigma}
(\hat{c}^{\phantom{\dagger}}_{\mathbf{i}\sigma})$ are creation
(destruction) operators for an electron at site $\mathbf{i}$ with spin
$\sigma$, $\mu$ is the chemical potential, and
$\hat{n}_{\mathbf{i}\sigma} =
\hat{c}^\dagger_{\mathbf{i}\sigma}\hat{c}^{\phantom{\dagger}}_{\mathbf{i}\sigma}$.
The first sum is taken over all nearest neighbor pairs $\langle
\mathbf{i, j} \rangle $ of a two-dimensional square lattice.  $t$ is
the nearest-neighbor hopping parameter which sets the energy scale
($t=1$), with the electronic bandwidth given by $W=8t$. At each site
are local harmonic oscillators of frequency $\omega_0$, with
independent degrees of freedom $\hat{X}_\mathbf{i} =
\sqrt{\frac{1}{2\omega_0}}\left(\hat{a}^\dagger_\mathbf{i} +
\hat{a}^{\phantom{\dagger}}_\mathbf{i} \right)$ and
$\hat{P}_\mathbf{i} =
\sqrt{\frac{\omega_0}{2}}\left(\hat{a}^\dagger_\mathbf{i} -
\hat{a}^{\phantom{\dagger}}_\mathbf{i} \right)$, where
$\hat{a}^\dagger_\mathbf{i}(\hat{a}_\mathbf{i})$ are phonon creation
(destruction) operators at site $\mathbf{i}$. The electron density
$\hat{n}_{\mathbf{i}\sigma}$ couples to the displacement
$\hat{X}_\mathbf{i}$ through a local electron-phonon coupling
$\lambda$. In this work we measure the electron-phonon coupling in
terms of the dimensionless quantity $\lambda_D = \lambda^2/\omega_0^2
\, W$.

We study the Holstein model using determinant quantum Monte Carlo
(DQMC) simulations \cite{blankenbecler81, santos03}. In DQMC, the
inverse temperature is expressed as $\beta = L_t \Delta\tau$, where
$L_t$ denotes the number of intervals along the imaginary time axis
with discretization $\Delta\tau$. The partition function $Z = \Tr
e^{-\beta \hat{H}} = \Tr e^{-\Delta\tau \hat{H}} e^{-\Delta\tau
  \hat{H}} \ldots e^{-\Delta\tau \hat{H}}$ can then be evaluated by
inserting complete sets of phonon position states $\ket{\{x_{i,
    \tau}\}}$ at each imaginary time slice.  Since the Hamiltonian is
quadratic in fermionic operators, these can be traced out, giving
\begin{equation}\label{partition}
Z = \int d\{x_{i,\tau}\} e^{-S_{Bose}} [\det(M(\{x_{i,\tau}\})]^2
\end{equation}
where
\begin{equation}\label{bose}
S_{Bose} = \Delta\tau \left[ \frac{\omega_0^2}{2} \sum_{\mathbf{i}, \tau}
 x_{\mathbf{i}, \tau}^2 + \sum_{\mathbf{i}, \tau} \left(\frac{x_{\mathbf{i}, \tau+1} - x_{\mathbf{i},
 \tau}}{\Delta\tau}\right)^2 \right].
\end{equation}

The harmonic oscillator terms in Eqn.~\eqref{ham} yield the `bosonic
action' term given by Eqn.~\eqref{bose}. The partition function also
includes the product of the determinant of two matrices
$M_\sigma(\{x_{\mathbf{i},\tau}\})$, one for each spin species $\sigma
= \{\uparrow, \downarrow\}$. These matrices depend on the phonon field
$\{ x_{\mathbf{i},\tau} \}$ only.  However, since $\hat X_\mathbf{i}$
couples in the same manner to the two species, the matrices $M_\sigma$
are identical, giving the square of a determinant.  An important
consequence is the absence of a sign problem at any electronic
filling. Physical quantities can be measured via Monte Carlo sampling
of the phonon field $\{ x_{\mathbf{i},\tau} \}$ and accumulating
appropriate combinations of the fermion Green's function ${\cal
  G}_{\mathbf{ij}} =\langle
c^{\phantom{\dagger}}_{\mathbf{i}\sigma}c^{\dagger}_{\mathbf{j}\sigma}
\rangle = [M^{-1}]_{\mathbf{ij}}$.  In our work we take $\Delta\tau =
0.125$. Trotter errors arising from the discretization of the
imaginary time axis are less than the statistical errors associated
with the Monte Carlo sampling for the charge and pair correlations
given below.

The electron-phonon coupling term gives rise to an effective
attractive electron-electron interaction $U_{\textrm{eff}} =
-\lambda^2/\omega_0^2$ which promotes the formation of local pairs. On
bipartite lattices this leads to CDW order at half-filling ($\langle
\hat{n}_{\mathbf{i}\uparrow} + \hat{n}_{\mathbf{i}\downarrow} \rangle
= 1$) with alternating doubly occupied and empty sites favored. This
occurs at $\mu=-\lambda^2/\omega_0^2$, which can be shown via a
particle-hole transformation. When the system is doped away from
half-filling, superconductivity can arise at sufficiently low
temperature due to the electron pairs becoming increasingly mobile.
In this work we study the competition between CDW and SC as electron
density is varied using DQMC, for a range of inverse temperatures
$\beta = T^{-1}$ as low as $\beta=28$. We fix $\lambda_D = 0.25$ and
study two fixed frequencies $\omega_0 = 1$ and $\omega_0 = 4$ for
lattices sizes with linear dimension up to $L=12$.

We characterize the nature of the charge ordering by calculating the
real-space, equal time, charge density correlation function
$C(\mathbf{r})$, given by 
\begin{equation}
C(\mathbf{r}) = \langle (\hat{n}_{\mathbf{i}\uparrow} +
\hat{n}_{\mathbf{i}\downarrow}) (\hat{n}_{\mathbf{i} +
  \mathbf{r}\uparrow} + \hat{n}_{\mathbf{i} + \mathbf{r}\downarrow})
\rangle,
\end{equation}
and its Fourier transform $S(\mathbf{q})$, the CDW structure factor
\begin{equation}
S(\mathbf{q}) = \frac{1}{N} \sum_{\mathbf{i},\mathbf{j}} 
e^{i\mathbf{q} \cdot (\mathbf{i}-\mathbf{j})} \langle \hat{n}_{\mathbf{i}} \hat{n}_{\mathbf{j}} \rangle.
\end{equation}
In the CDW ordered phase, $C(\mathbf{r})$ becomes long ranged and
$S(\mathbf{q})$ grows in proportion to the lattice size $N=L^2$ at the
appropriate ordering wavevector $\mathbf{q}=(q_x, q_y)$. In the
absence of CDW order, the charge density correlations are short ranged
and $S(\mathbf{q})$ should exhibit no lattice size dependence.  The
superconducting response of the system is analyzed by the s-wave pair
susceptibility
\begin{equation}
P_s = \frac{1}{N} \int_0^\beta \langle \Delta(\tau) \Delta^\dagger(0) \rangle d\tau,
\end{equation}
where $\Delta(\tau) = \sum_\mathbf{i} c_{\mathbf{i}\downarrow}(\tau)
c_{\mathbf{i}\uparrow}(\tau)$.  Similarly, an enhancement in the pair
susceptibility and the observation of lattice size dependence in $P_s$
as the temperature is lowered can be used to detect the onset of SC
order.  We use the susceptibility to study SC, as opposed to an equal
time structure factor, because it provides a more robust signal which
is useful for exploring off-diagonal long range order of the KT type.

\begin{figure}[t!]
\centering
\includegraphics[width=\columnwidth]{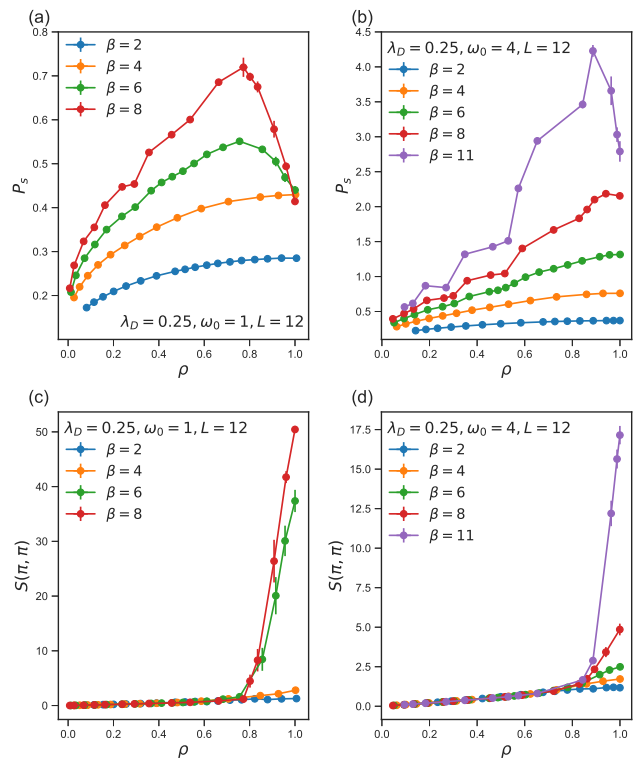}
\caption{\footnotesize{(a) S-wave pair susceptibility $P_s$ as a
    function of electron density $\rho$ for $\omega_0=1$ and
    $\lambda_D=0.25$. (b) $P_s$ vs.~$\rho$ for $\omega_0=4$ and
    $\lambda_D=0.25$. (c) CDW structure factor $S(\pi, \pi)$ as a
    function of electron density $\rho$ for $\omega_0=1$ and
    $\lambda_D=0.25$. (d) $S(\pi, \pi)$ vs.~$\rho$ for $\omega_0=4$
    and $\lambda_D=0.25$.  Data are shown for a $12 \times 12$ lattice
    for inverse temperatures $\beta = 2$, $4$, $6$, $8$ and $11$.}}
    \label{fig:sweeps}
\end{figure}

\begin{figure}[ht!]
\centering
\includegraphics[width=\columnwidth]{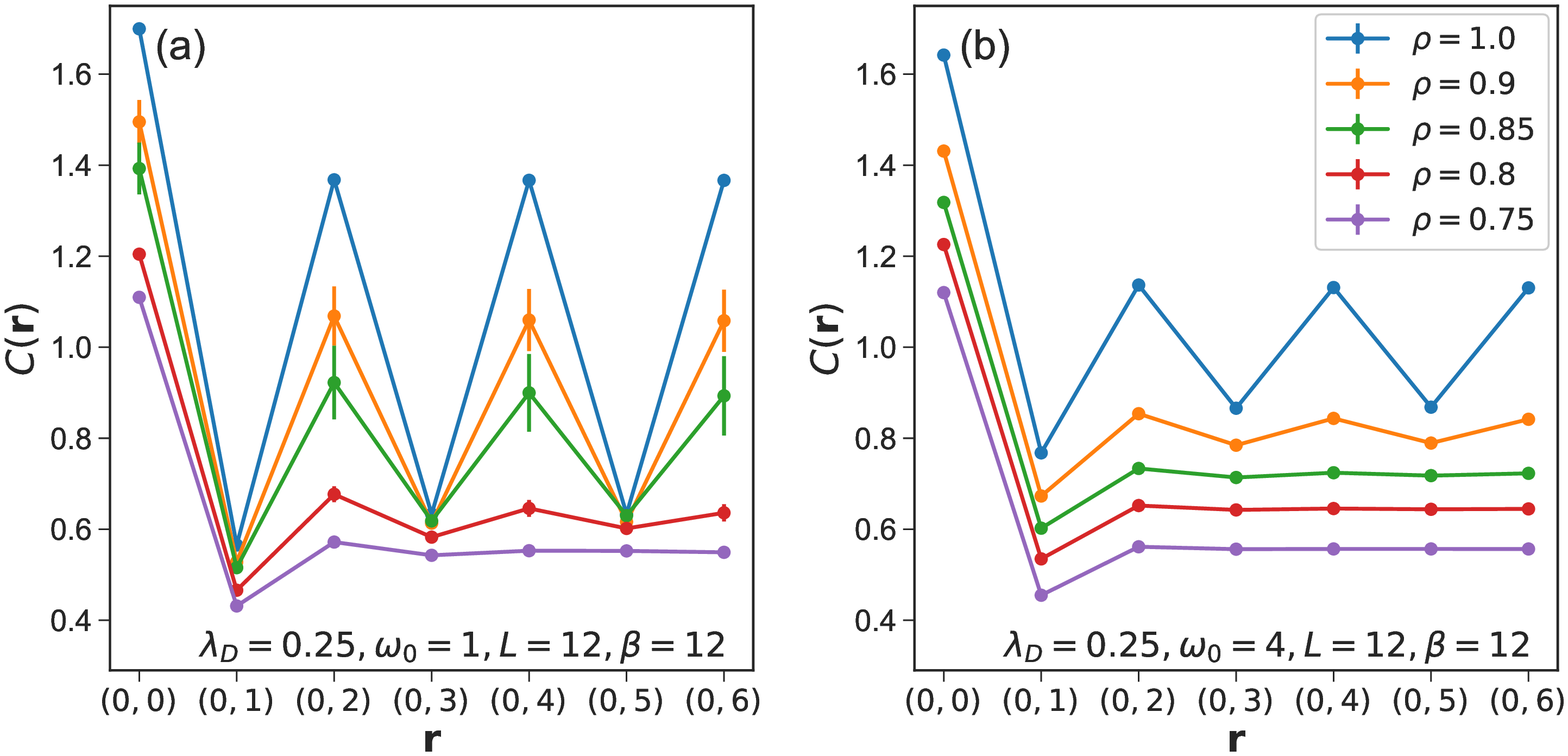}
\caption{\footnotesize{(a) Charge density correlation function
    $C(\mathbf{r})$ as a function of site separation $\mathbf{r}$, for
    a $12 \times 12$ lattice at $\beta=12$, with
    $\mathbf{r}=(0,1)-(0,6)$ in units of the lattice spacing.  Results
    are shown for $\omega_0=1$ and $\lambda_D=0.25$ for fixed electron
    densities: $\rho = 1$, $0.9$, $0.85$, $0.8$ and $0.75$. (b)
    $C(\mathbf{r})$ vs.~$\mathbf{r}$ for $\omega_0=4$ and
    $\lambda_D=0.25$.}}
    \label{fig:corr}
\end{figure}

\begin{figure*}[]
\centering
\includegraphics[width=\textwidth]{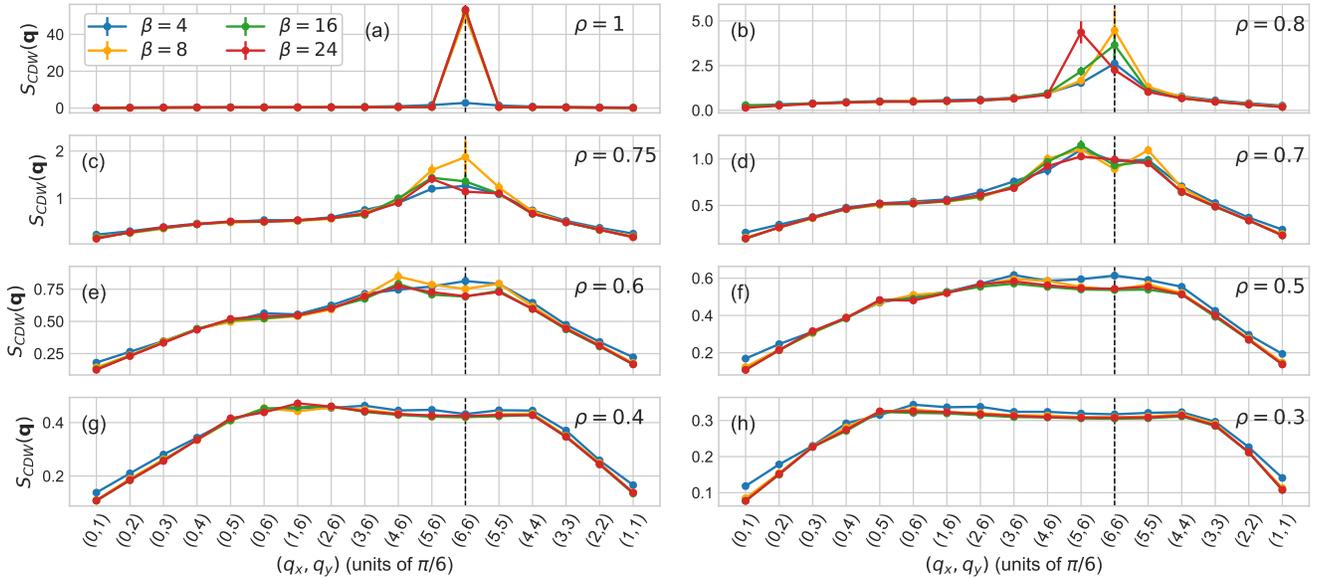}
\caption{\footnotesize{Variation of $S(\mathbf{q})$ with wavevector
    $\mathbf{q}$ for a $12 \times 12$ lattice, for $\omega_0=1$,
    $\lambda_D=0.25$. A triangular path through the Brillouin zone is
    taken from $\mathbf{q} = (0, \pi/6)$ to $(0, \pi)$ to $(\pi, \pi)$
    to $(\pi/6, \pi/6)$. Results are shown for inverse temperatures
    $\beta=4, 8, 16$ and $24$ for electron densities in the range
    $\rho=0.3-1.0$, specified in the upper-right corner of plots
    (a)--(h). In each plot the dashed line indicates the location of
    $\mathbf{q} = (\pi, \pi)$.
    }
    }
    \label{fig:Sqw1}
\end{figure*}

\section{III. Results and Discussion}
At half-filling, 
i.e.~$\rho = \langle \hat{n}_{\mathbf{i}\uparrow} + \hat{n}_{\mathbf{i}\downarrow} \rangle = 1$, 
it is known that checkerboard CDW order dominates on the square lattice with ordering wavevector
$\mathbf{q}=(\pi, \pi)$. This occurs above the inverse critical temperature
$\beta_{\rm cdw} = 6.0 \pm 0.1$ for $\omega_0 = 1$ and $\beta_{\rm
  cdw} \approx 13$ for $\omega_0 = 4$, with $\lambda_D = 0.25$ in both
cases \cite{costa18}. By varying the chemical potential, we dope the
system away from half-filling and study the behavior of both $S(\pi,
\pi)$ and $P_s$ as a function of electron density, as shown in
Figs.~\ref{fig:sweeps}(a)--(d) for $\omega_0 = 1$ and $\omega_0 = 4$ at $\lambda_D =
0.25$.  In both cases, $S(\pi, \pi)$ is significantly enhanced at
$\rho = 1$ when the inverse temperature approaches $\beta_{\rm cdw}$,
but rapidly falls off when doped away from half-filling, and is highly
suppressed below $\rho \approx 0.75$ for $\omega_0 = 1$.
Simultaneously, the s-wave pair susceptibility becomes enhanced away
from half-filling, reaching a maximum within the density range $\rho =
0.6 \textrm{--} 0.7$. When the phonon frequency is increased to
$\omega_0 = 4$, $P_s$ increases in magnitude, while $S(\pi, \pi)$ is
diminished and becomes highly suppressed at a density closer to
half-filling, at approximately $\rho \approx 0.85$.

The CDW ordering which occurs at half-filling above $\beta_{\rm cdw}$
on the square lattice is a checkerboard pattern of alternating doubly
occupied and empty sites. This becomes evident by plotting the
real-space charge density correlation function $C(\mathbf{r})$ against
site separation, as shown in Fig.~\ref{fig:corr} for a $12\times12$ lattice at
$\beta=12$, for (a) $\omega_0 = 1$ and (b) $\omega_0 = 4$. The
alternating high and low correlations at $\rho=1$ are smoothed out as
the density is lowered, with $C(\mathbf{r})$ becoming flat around
$\rho \lesssim 0.75$ for $\omega_0 = 1$ and $\rho \lesssim 0.85$ for
$\omega_0 = 4$. Increasing the phonon frequency inhibits CDW order,
which is reflected by the smaller charge density correlations (at
$\beta = 12$) for $\omega_0 = 4$, and the fact that the alternating CDW
pattern is more rapidly suppressed for this frequency when doped away
from half-filling.
 
 \begin{figure*}[t!]
\centering
\includegraphics[width=\textwidth]{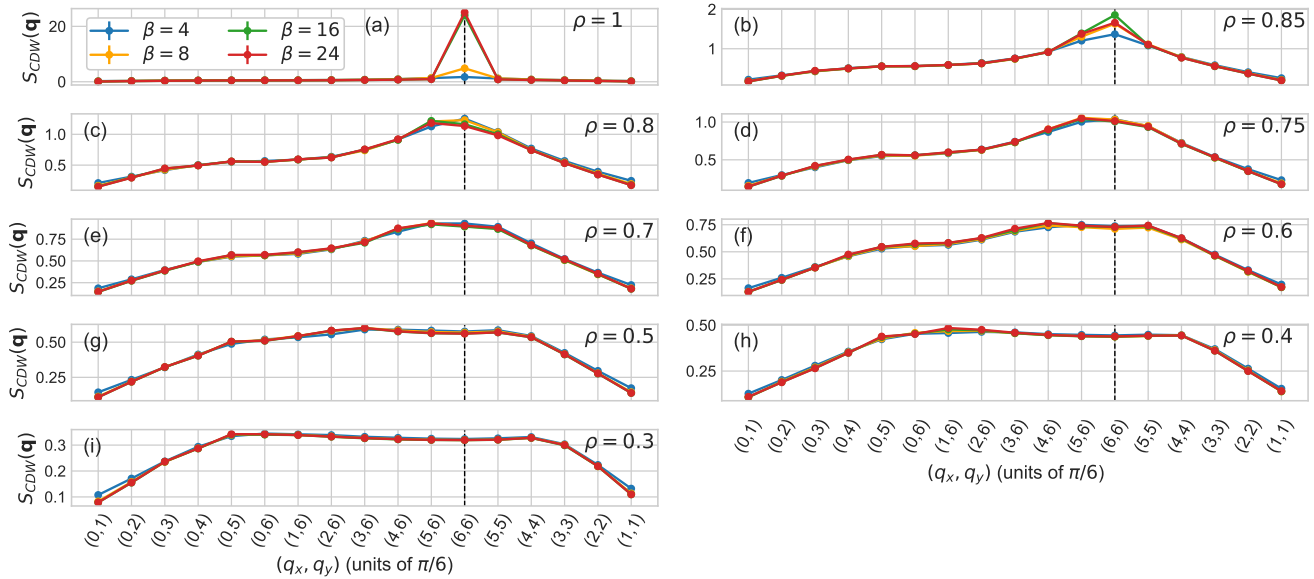}
\caption{\footnotesize{Variation of $S(\mathbf{q})$ with wavevector
    $\mathbf{q}$ for a $12 \times 12$ lattice, for $\omega_0=4$,
    $\lambda_D=0.25$. A triangular path through the Brillouin zone is
    taken from $\mathbf{q} = (0, \pi/6)$ to $(0, \pi)$ to $(\pi, \pi)$
    to $(\pi/6, \pi/6)$. Results are shown for inverse temperatures
    $\beta=4, 8, 16$ and $24$ for electron densities in the range
    $\rho=0.3-1.0$, specified in the upper-right corner of plots
    (a)--(i). In each plot the dashed line indicates the location of
    $\mathbf{q} = (\pi, \pi)$.} 
    }
    \label{fig:Sqw4}
\end{figure*}

\begin{figure}[t!]
\centering
\includegraphics[width=\columnwidth]{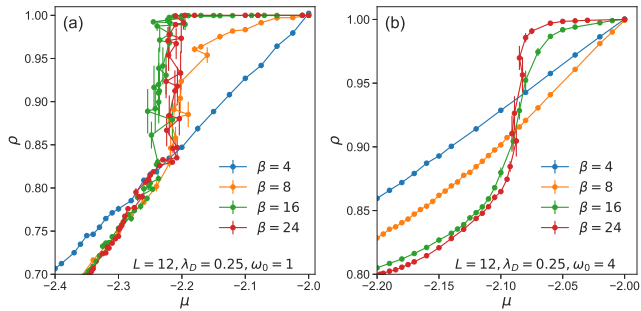}
\caption{\footnotesize{ Density $\rho$ as a function of chemical
    potential $\mu$ approaching the CDW transition at
    half-filling. Results shown for $L = 12$ lattices with
    $\lambda_D=0.25$, for phonon frequencies (a) $\omega_0=1$ and (b)
    $\omega_0=4$. The data suggest a discontinuous jump prior to entry
    to the incompressible CDW region.  }}
    \label{fig:rhovsmu}
\end{figure}

At half-filling the square lattice exhibits perfect Fermi surface
nesting (FSN) at $\mathbf{q}=(\pi, \pi)$ in the absence of any
next-nearest neighbor hopping term, resulting in a peak in
$S(\mathbf{q})$ at this wavevector. However when doped away from
half-filling, the Fermi surface becomes distorted and perfect FSN no
longer occurs.  In Figs.~\ref{fig:Sqw1}(a)--(h) we show the variation of
$S(\mathbf{q})$ with wavevector $\mathbf{q}=(q_x, q_y)$, taken on a
triangular path through the Brillouin zone, for a $12 \times 12$
lattice at $\beta=4$, $8$, $16$ and $24$, for $\lambda_D = 0.25$,
$\omega_0 = 1$. Results are shown for a range of electron densities
from $\rho = 0.3 \textrm{--} 1.0$. $S({\bf q})$ is not shown for small dopings away from half-filling.
This will be further discussed in the interpretation of $\rho(\mu)$ shown
in Fig.~\ref{fig:rhovsmu}. Away from half-filling the peak
magnitude of $S(\pi, \pi)$ is rapidly suppressed, reduced by a factor
of 10 by $\rho\approx0.8$, and falling by another order of magnitude
by $\rho\approx0.5$ (note the vertical scale of each plot).

  There is an important comment to make concerning the behavior at $\rho \approx 0.8$, where the location of the peak appears to shift to the
  wavevector nearest to $(\pi, \pi)$, i.e.~$\mathbf{q}=(5\pi/6, \pi)$
  as shown in Fig.~\ref{fig:Sqw1}(d), with the shift occurring at low temperature
  ($\beta \approx 24$). The magnitude of $S(5\pi/6, \pi)$ at $\rho
  \approx 0.8$ grows as the temperature is lowered, becoming
  substantially enhanced at $\beta=24$. Although this suggests the 
  {\it possible}
  existence of an incommensurate CDW phase at $\rho\approx0.8$, the rather coarse 
  discrete momentum grid $q=\frac{2 \pi}{L}\{0,1,\cdots L\}$ precludes any conclusive
  statement.
  
When the system is doped even further from half-filling, as
in Figs.~\ref{fig:Sqw1}(c)--(h), we do not observe any significant enhancement in
$S(\mathbf{q})$ at any wavevector as the temperature is lowered from
$\beta=4 $ to $\beta=24$. The magnitude of $S(\mathbf{q})$ remains
approximately constant over this temperature range for all values of
$\mathbf{q}$, as shown in Figs.~\ref{fig:Sqw1}(c)--(h) for $\rho\leq0.75$. In
particular, within the density range $\rho = 0.6 \textrm{--} 0.7$, for
which we observe a peak in the s-wave pair susceptibility, we find no
indication of a coexisting CDW phase for any ordering wavevector.

Increasing the phonon frequency to $\omega_0=4$, we find qualitatively
similar results as shown in Figs.~\ref{fig:Sqw4}(a)--(i), however there is no
indication of CDW ordering at any particular wavevector for any
electron density, other than at $\mathbf{q}=(\pi, \pi)$ at low
temperature. The magnitude of $S(\pi, \pi)$ near half filling is also
considerably suppressed compared to $\omega_0=1$, which is expected
since increasing the phonon frequency inhibits CDW
order. Although the peak in $S(\mathbf{q})$ shifts to $\mathbf{q}=(5\pi/6, \pi)$ at $\rho \approx 0.8$ as the temperature is reduced, there is no significant enhancement in the magnitude of $S(\mathbf{q})$ at this wavevector as temperature is lowered from $\beta=4$ to $\beta=24$, in contrast to the behavior at $\omega_0=1$ (Fig.~\ref{fig:Sqw1}).

\begin{figure}[t!]
\centering
\includegraphics[width=\columnwidth]{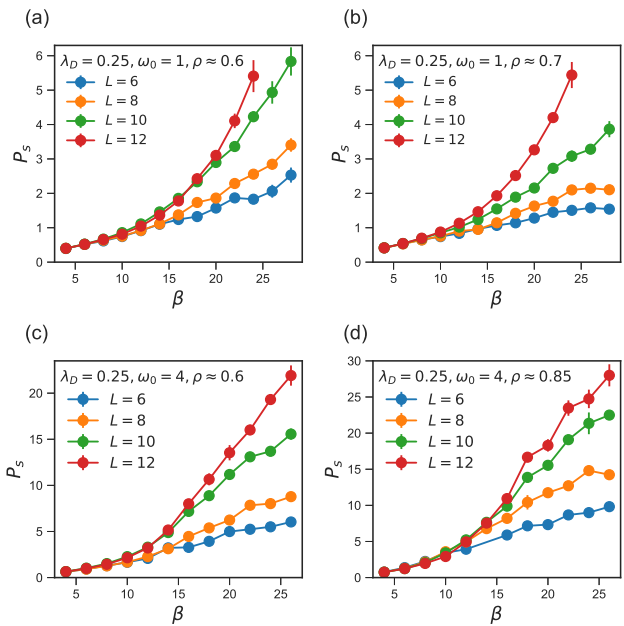}
\caption{\footnotesize{S-wave pair susceptibility as a function of inverse temperature $\beta$
 for lattice sizes of linear dimension $L = 6$, $8$, $10$ and $12$, for the four fixed
 densities studied: (a) $\rho = 0.6$ and (b) $\rho = 0.7$ for $\lambda_D=0.25$, $\omega_0=1$.
 For increased phonon frequency $\omega_0=4$, we fix (c) $\rho = 0.6$ and (d) $\rho = 0.85$ 
 with the same dimensionless coupling $\lambda_D=0.25$ and (a,b).}}
 \label{fig:Psbeta}
\end{figure}

In order to determine the critical inverse temperature $\beta_{\rm
  sc}$ for the SC transition, we first tune the chemical potential to
achieve a fixed target density and study $P_s$ as a function of
$\beta$, for several different lattice sizes. Since $P_s$ appears to
peak in the range $\rho=0.6\textrm{--}0.7$ for $\omega_0 = 1$,
$\lambda_D = 0.25$, we choose to study two fixed densities $\rho =
0.6$ and $\rho = 0.7$ for this phonon frequency. For $\omega_0 = 4$,
$\lambda_D=0.25$, since CDW correlations appear highly suppressed
closer to half-filling, we fix $\rho=0.85$ and also study $\rho=0.6$
for comparison. In Figs.~\ref{fig:Psbeta}(a)--(d) we show $P_s(\beta)$ for lattices
of linear dimension $L=6, 8, 10$ and $12$ for these four parameter
sets. For each case, we find at low $\beta$ (high $T$), $P_s$ is
relatively small and is independent of lattice size, however as the
temperature is lowered, $P_s$ grows and becomes dependent on $L$. This
suggests the onset of the SC phase, because when correlations become
long range they will be sensitive to the lattice size for a finite
system. We can therefore apply a finite-size scaling analysis to
confirm the existence of a critical inverse temperature $\beta_{\rm
  sc}$ for the SC transition, and determine its value.

\begin{figure}[ht!]
\centering
\includegraphics[width=\columnwidth]{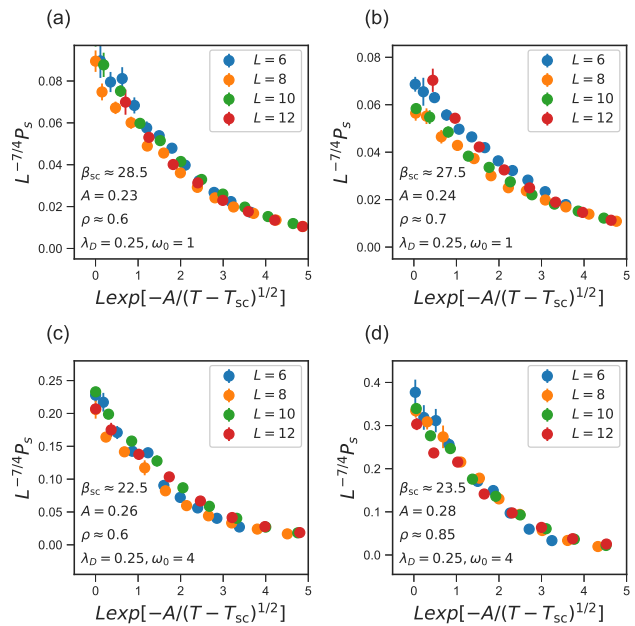}
\caption{\footnotesize{Finite size scaling of the s-wave pair susceptibility data obtained for
 the four parameter sets shown in Figs.~\ref{fig:Psbeta}(a)--(d). The critical inverse temperature $\beta_{\rm sc}$
 and scaling factor $A$ which yields the best data collapse is indicated in the inset of each plot. }}
 \label{fig:collapse}
\end{figure}

In the two dimensional superconducting transition, the order parameter
possesses $U(1)$ gauge symmetry and thus the universality class is the
same as the 2D XY model. Hence we expect a Kosterlitz-Thouless (KT)
transition to a quasi-long-range ordered phase, for which the critical
exponents and scaling behavior of the order parameter are known
\cite{kosterlitz74}. For a finite-size system of linear dimension $L$,
we have that
\begin{equation}
P_s = L^{2-\eta} f\left(\frac{L}{\xi}\right)
\end{equation}
with $\eta = 1/4$, and as $T \to T_{\rm sc}^+$ the correlation length $\xi$ scales as
\begin{equation}
\xi \sim \exp\left[A \left( T - T_{\rm sc} \right)^{-1/2}\right] 
\end{equation}
where $A$ is a constant and $T_{\rm sc}$ is the critical temperature. Therefore near $T_{\rm sc}$,
 plotting $P_s L^{-7/4}$ as a function of $L \exp[-A\left(T-T_{\rm sc}\right)^{-1/2}]$ for a
 range of lattice sizes should result in a data collapse onto a single universal curve, as shown
 in Figs.~\ref{fig:collapse}(a)--(d) for the four parameter sets studied. For $\lambda_D=0.25$, $\omega_0=1$,
 we find the best data collapse occurs at $\beta_{\rm sc} \approx 28.5 \pm 1.0$ for $\rho=0.6$ and
  $\beta_{\rm sc} \approx 27.5 \pm 1.0$ for $\rho=0.7$. Keeping the dimensionless electron-phonon 
coupling fixed at $\lambda_D=0.25$, increasing phonon frequency to $\omega_0=4$ raises
 the critical temperature, and we find the best data collapse at $\beta_{\rm sc} \approx 22.5 \pm 1.0$
 for $\rho=0.6$ and $\beta_{\rm sc} \approx 23.5 \pm 1.0$ for $\rho=0.85$. Our value of $\beta_{\rm sc}$ 
for $\omega_0 = 1$ lies slightly below the range of $\beta_{\rm sc} = 30 \textrm{--} 40$
 suggested by Vekic et al \cite{vekic92}, although their estimate was performed
 using data rather far from the scaling region.  Meanwhile, our estimate
of $\beta_{\rm sc}$ for $\omega_0 = 4$ at $\rho=0.85$ is higher than the previous 
 $\beta_{\rm sc} \approx 12$ at $\rho = 0.8$. 
The larger values of $L$ and $\beta$ accessed in this
study allow a more robust finite-size scaling for the KT transition. We also note that for the lower phonon frequency we study, for which the ME approximation would be more justifiable than for $\omega_0=4$, recent ME calculations \cite{kivelson_private} have estimated $T_{\rm sc}$ for the parameters shown in Fig.~\ref{fig:Psbeta}(a), yielding a value within approximately $10\%$ our result.

We note that increasing phonon frequency simultaneously raises $T_{\rm
  sc}$ for the SC transition, and lowers $T_{\rm cdw}$ for the CDW
transition at half-filling (from $T_{\rm cdw} \approx t/6$ at
$\omega_0/t=1$ to $T_{\rm cdw} \approx t/13$ at $\omega_0/t = 4$
\cite{costa18}), illustrating the competition between SC and CDW order
in the Holstein model. This is as expected since as $\omega_0$ is
lowered, the harmonic oscillators on each site become more classical,
reducing quantum fluctuations. As a result, bipolarons localize more
readily, enhancing CDW order \cite{li19}. Conversely, it is known that
in the anti-adiabatic limit ($\omega_0 \to \infty$) the Holstein model
can be mapped onto the attractive Hubbard model \cite{scalettar89,
  noack91, vekic92} with
$U_{\textrm{eff}}=-\lambda^2/\omega_0^2=-\lambda_D W$ \cite{berger95},
which has been shown to possess a finite temperature superconducting
KT transition away from half-filling \cite{scalettarNegU89, moreo91,
  paiva04}. Thus one expects SC correlations to be enhanced in the
Holstein model at larger values of $\omega_0$, as we have confirmed
here. Furthermore, in the attractive Hubbard model, the SC and CDW
correlations are degenerate at half-filling, leading to a continuous
order parameter in the Heisenberg universality class and the absence
of a finite-temperature transition (i.e.~$T_c = 0$) in 2D. At
half-filling, the CDW order parameter $S(\pi, \pi)$ is therefore
reduced, with $P_s$ increasing simultaneously in the limit $T \to 0$.
We thus expect similar behavior in the Holstein model as $\omega_0 \to
\infty$, which we have observed as an enhancement in $P_s$ and a
reduction in $S(\pi, \pi)$ at $\omega_0 = 4$ at half-filling, as shown
in Figs.~\ref{fig:sweeps}(a)--(d). We also note that studies of the attractive
Hubbard model have found $T_{\rm sc}$ is maximal at around $U/t
\approx -5$, for which $T_{\rm sc}/t\approx 0.15$ occurs at a filling
$\rho=0.7$ \cite{paiva10}. Since this effective coupling corresponds
to a larger $\lambda_D$ value than we study in this work, this
suggests raising $\lambda_D$ could enhance $T_{\rm sc}$ at large
phonon frequencies. We have determined $T_{\rm sc}$ values for
$-\lambda_D W = -2$ in this work, which one can compare to recent
estimates of $T_{\rm sc}$ in the attractive Hubbard model
\cite{costa_private}: for $U=-2.0$, $\beta_{\rm sc} = 19.0$ at
$\rho=0.7$, and $\beta_{\rm sc} = 13.5$ at $\rho=0.87$, while for
$U=-2.5$, $\beta_{\rm sc} = 23.0$ at $\rho=0.5$. However, for
$\omega_0=1$ and $\omega_0=4$, the actual on-site interaction will be
smaller than in the anti-adiabatic limit (i.e.~$|U|<2$), giving a
lower $T_{\rm sc}$, and the attractive Hubbard model thus provides an
upper bound on $T_{\rm sc}$ in the Holstein model. Our estimates of
$T_{\rm sc}$ at $\omega_0=1$ and $\omega_0=4$ are therefore quite
consistent with those of the attractive Hubbard model.

We conclude the presentation of our results by noting that $\rho(\mu)$
appears to exhibit a discontinuous jump approaching the CDW transition
at half-filling, as shown in Figs.~\ref{fig:rhovsmu}(a) and (b) for $\omega_0=1$ and
$\omega_0=4$. In both cases, we have that half-filling ($\rho=1$)
occurs at a chemical potential of
$\mu=-\lambda^2/\omega_0^2=-2$. Below $T_{\rm cdw}$, the formation of
a plateau at $\rho=1$ indicates the opening of the CDW gap. However,
well below the transition temperature ($\beta_{\rm cdw} = 6.0 \pm 0.1$
for $\omega_0=1$ and $\beta_{\rm cdw} \approx 13$ for $\omega_0=4$) we
observe a discontinuous jump in electron density as the chemical
potential is varied, occurring for $\rho \gtrsim 0.8$ for
$\omega_0=1$, and $\rho \gtrsim 0.9$ for $\omega_0=4$ (with
$\lambda_D=0.25$ in both cases). We note that these density ranges
correspond roughly to the regions over which $S(\pi, \pi)$ grows
rapidly, occurring closer to half-filling for greater $\omega_0$, as
shown previously in Figs.~\ref{fig:sweeps}(c) and (d). The jump is less abrupt for $\omega_0=4$ but
becomes apparent at $\beta=24$, whereas a clear discontinuity emerges
for $\beta \geq 16$ for $\omega_0=1$. 
This indicates
finite temperature fluctuations smooth the jump more at higher frequencies.

In both cases, the jump is
accompanied by an increase in the error in $\rho$ for data close to
half-filling, possibly indicating fluctuations of the system between
densities on either side of the discontinuity. This discontinuity may
be related to the zero temperature transition from SC to commensurate
CDW order, which has been observed to be first order
\cite{esterlis19}.

\section{IV. Summary and Conclusions}

In previous QMC studies, the CDW transition temperature $T_{\rm cdw}$
of the Holstein model at half-filling has been determined for various
two-dimensional systems, including the square, honeycomb, and Lieb
lattices. However, the superconducting transition away from
half-filling in the square lattice has been much less well
characterized, since it occurs at challengingly large values of the
inverse temperature $\beta$ as well as scaling in the spatial lattice
size $L$. Moreover, away from half-filling, no analytical expression
for $\rho(\mu)$ can be used to achieve a fixed target density
\cite{miles20}, necessitating a cumbersome tuning of $\mu$ for each
lattice size and $\beta$. In this work, we have studied larger systems
(up to $L = 12$) and lower temperatures (up to $\beta=28)$ than in
previous work, and have determined several estimates of $T_{\rm sc}$
for various electron densities (fixed via tuning the chemical
potential) and phonon frequencies $\omega_0$, through a finite-size
scaling analysis of pair susceptibility. We observe the onset of SC at
temperatures $T_{\rm sc} \lesssim W/160$ in each case studied. Here
$W=8\,t$ is the non-interacting bandwidth and $t$ is the nearest
neighbor hopping amplitude.

Specifically, for dimensionless electron-phonon coupling
$\lambda_D=0.25$, and phonon frequency $\omega_0/t=1$, we estimate
$T_{\rm sc} \approx W/228 = t/28.5$ for $\rho = 0.6$ and $T_{\rm sc}
\approx W/220 = t/27.5$ for $\rho = 0.7$.  For $\lambda_D=0.25$,
$\omega_0=4$, we estimate $T_{\rm sc} \approx W/180 = t/22.5$ for
$\rho = 0.6$ and $T_{\rm sc} \approx W/228 = t/23.5$ for $\rho =
0.85$.

Several features illustrating the competition between CDW order and SC
in the doped Holstein model emerge from our analysis. In particular,
the strong checkerboard CDW order present at half-filling below
$T_{\rm cdw}$ (corresponding to a peak in $S(\pi, \pi)$) is rapidly
suppressed as the system is doped, with SC correlations becoming
maximal in the region $\rho = 0.6\textrm{--}0.7$ for $\lambda_D=0.25$,
$\omega_0=1$. However, at an intermediate electron density of
  approximately $\rho \approx 0.8$, we observe evidence of a {\it possible}
  incommensurate CDW phase, with the peak in $S(\mathbf{q})$ shifting
  slightly from $\mathbf{q}=(\pi, \pi)$ to $\mathbf{q}=(5\pi/6, \pi)$
  at low temperature.  Definitive analysis of this point is precluded by
  the finite momentum grids currently accessible to present QMC capabilities. 
 No evidence of a distinctly different kind of
charge ordering (e.g.~stripe order) is observed away from
half-filling.

It is interesting to note that our estimates of $T_{\rm sc}$ in the
doped Holstein model are similar in magnitude to $T_{\rm sc}$ in the
half-filled case with non-zero phonon dispersion $\Delta \omega
/\omega_0 = 0.1$, where DQMC simulations \cite{costa18} have
determined $T_{\rm sc} \approx t/26$ for $\lambda_D=0.25$,
$\omega_0=4$. Further, it has been proposed \cite{esterlis18_2} that
an upper bound on $T_{\rm sc}$ exists which is $T_{\rm sc} \lesssim
\bar{\omega}/10$, where $\bar{\omega}\leq\omega_0$ is a characteristic
phonon frequency no larger than the bare phonon frequency, and that
for an optimal value of $\lambda_D$, $T_{\rm sc}$ should roughly
saturate at this value. Since our estimates of $T_{\rm sc}$ lie below
this upper bound, this suggests it may be possible to increase the
transition temperature by increasing $\lambda_D$. Recently, a QMC
method based on Langevin updates of the phonon degrees of freedom
\cite{batrouni19, batrounicomp19} has also made studies of the cubic
Holstein model amenable to simulation, and it has been found that
$T_{\rm cdw}$ at half-filling is increased roughly by a factor of two
compared to various two-dimensional geometries \cite{cohenstead20}.
We anticipate that in future studies of the 3D Holstein model one
might similarly expect higher values of $T_{\rm sc}$ away from
half-filling, since the model will exhibit a more robust transition to
long-ranged superconducting order, in contrast with the KT transition
in two dimensions observed in this work. \newline

\section{Acknowledgments} 
We would like to thank Steven Kivelson, Ilya Esterlis, and Natanael
Costa for insightful comments on this work. The work of O.B.~and
R.S.~was supported by the grant DE‐SC0014671 funded by the
U.S. Department of Energy, Office of Science.

\bibliography{bradley}

\end{document}